\documentclass[twocolumn]{aastex7}

\usepackage{amsmath}
\usepackage{microtype}
\usepackage{multirow}
\usepackage{graphicx}
\usepackage{flafter}
\usepackage{booktabs}
\usepackage{hyperref}
\usepackage{cleveref}

\makeatletter
\newenvironment{compactwidetext}{%
\if@two@col
  \par\ignorespaces
  \onecolumngrid
  \vspace{-0.45\baselineskip}%
  \noindent\rule{\textwidth}{0.35pt}\par
  \vspace{0.20\baselineskip}%
  \prep@math@patch
\fi}{%
\if@two@col
  \par\vspace{-0.35\baselineskip}%
  \noindent\rule{\textwidth}{0.35pt}\par
  \vspace{0.35\baselineskip}%
  \twocolumngrid\global\@ignoretrue
  \@endpetrue
\fi}
\makeatother
\providecommand{\Letter}{\ensuremath{\ast}}

\newcommand{\code}[1]{\texttt{#1}}

\def\ba{\begin{eqnarray}}
\def\ea{\end{eqnarray}}

\begin{document}

\title{Cosmological inference from the eBOSS QSO full-shape analysis with optimal redshift weights}

\author[0000-0002-2671-9078]{Xiaoyong Mu}
\affiliation{National Astronomical Observatories, Chinese Academy of Sciences, Beijing, 100101, P.R.China}
\affiliation{School of Astronomy and Space Sciences, University of Chinese Academy of Sciences, Beijing, 100049, P.R.China}
\email{mouxiaoyong15@mails.ucas.ac.cn}  

\author{Zhuo-Heng Li}
\affiliation{National Astronomical Observatories, Chinese Academy of Sciences, Beijing, 100101, P.R.China}
\affiliation{School of Astronomy and Space Sciences, University of Chinese Academy of Sciences, Beijing, 100049, P.R.China}
\email{lizhuoheng20@mails.ucas.ac.cn}

\author{Wentao Luo}
\affiliation{National Astronomical Observatories, Chinese Academy of Sciences, Beijing, 100101, P.R.China}
\affiliation{School of Astronomy and Space Sciences, University of Chinese Academy of Sciences, Beijing, 100049, P.R.China}
\email{luowentao20@mails.ucas.ac.cn}

\author[0000-0001-7756-8479]{Yuting Wang\textsuperscript{\Letter}}
\affiliation{National Astronomical Observatories, Chinese Academy of Sciences, Beijing, 100101, P.R.China}
\email{ytwang@nao.cas.cn} 

\correspondingauthor{Yuting Wang, Gong-Bo Zhao (\url{ytwang@nao.cas.cn, gbzhao@nao.cas.cn})}
\author[0000-0003-4726-6714]{Gong-Bo Zhao\textsuperscript{\Letter}}
\affiliation{National Astronomical Observatories, Chinese Academy of Sciences, Beijing, 100101, P.R.China}
\affiliation{School of Astronomy and Space Sciences, University of Chinese Academy of Sciences, Beijing, 100049, P.R.China}
\affiliation{Institute for Frontiers in Astronomy and Astrophysics, Beijing Normal University, Beijing, 102206, P.R.China}
\email{gbzhao@nao.cas.cn} 

\begin{abstract}
We present a full-shape power-spectrum analysis of the eBOSS DR16 quasar sample with optimal redshift weights. The DR16 QSO catalog contains 343,708 quasars over $0.8<z<2.2$, a redshift interval broad enough to contain useful light-cone evolution but not naturally captured by a single effective-redshift measurement. We construct Karhunen--Lo\`eve weights for the parameters of interest and measure the resulting monopole and quadrupole with a cross-correlation estimator, which remains well defined for sign-changing weights. The theoretical spectra are convolved with the measured Fourier-space survey-window kernels for each Galactic cap and weighting scheme, and both the covariance matrix and the end-to-end validation are based on 1000 EZ light-cone mock catalogs. In $\Lambda$CDM, the redshift-weighted and standard analyses give consistent constraints, as expected from the near-standard effective redshifts of the weights targeting $h$, $\Omega_{\rm m}$, and $A_s$. In the Chevallier--Polarski--Linder (CPL) model, the redshift-weighted DR16 analysis reduces the marginalized uncertainties on $H_0$, $\sigma_8$, and $w_0$ by $43.3\%$, $19.7\%$, and $20.5\%$, respectively, and turns the standard one-sided constraint on $w_a$ into a bounded posterior, $w_a=-0.98^{+1.0}_{-1.3}$. The gain is therefore concentrated where the model contains genuine redshift evolution, demonstrating that optimal redshift weighting can recover tomographic information from a wide QSO light cone while keeping the full-shape data vector compact.
\end{abstract}


\section{Introduction}\label{sec:intro}

Large-scale structure surveys constrain the expansion history, the growth of structure, and the shape of the matter power spectrum. The baryon acoustic oscillation (BAO) feature provides a standard ruler for distance measurements \citep{Eisenstein2005,Cole2005}, while redshift-space distortions (RSD) probe peculiar velocities and hence the growth rate \citep{Kaiser1987,Peacock2001}. Full-shape analyses retain the broadband clustering information in addition to the compressed BAO and RSD summaries, and have become a useful route for extracting cosmological information from spectroscopic surveys \citep{Neveux2020,eBOSS:2020yzd}.

The eBOSS DR16 QSO sample is well suited to this problem because it spans $0.8\lesssim z\lesssim2.2$ \citep{Dawson2016,Ross2020}. This wide redshift range gives access to the evolution of geometry, growth, bias, and number density, but it also creates a practical choice. A single effective-redshift measurement is compact but averages over the light cone; a multi-bin tomographic analysis keeps more redshift information but requires several correlated clustering measurements and a larger covariance matrix.

Optimal redshift weighting offers a middle ground. The survey is kept as a single sample, while each object is assigned redshift weights chosen to preserve the Fisher information on selected parameters \citep{Tegmark1997,Zhu:2014ica}. This idea has been applied to BAO and RSD measurements in configuration and Fourier space \citep{Ruggeri2017,Wang2018,eBOSS:2018yfg,Ruggeri2019}. Here we extend it to the full-shape QSO power spectrum, where the broadband signal may respond differently to redshift-dependent parameters than to the standard effective-redshift combination.

This analysis tests whether optimal weights can recover the redshift-evolution information in the QSO light cone without introducing the larger data vector and covariance matrix of a tomographic full-shape measurement. We model the monopole and quadrupole with the EFTofLSS, measure the weighted spectra using a cross-correlation estimator that also works for sign-changing weights, and compare standard and weighted constraints in both $\Lambda$CDM and the CPL dark-energy model. The pipeline is validated on 1000 EZ light-cone mocks before being applied to the DR16 QSO data.

The remainder of the paper is organized as follows. Section~\ref{sec:method} summarizes the full-shape model and the KL compression. Section~\ref{sec:weights} describes the data, mocks, redshift weights, power-spectrum measurements, Fourier-space window convolution, and likelihood. Section~\ref{sec:result} presents the mock validation and the DR16 constraints. Section~\ref{sec:conclu} gives the conclusions. Appendix~\ref{app:cross_auto} compares the cross-correlation estimator with the conventional auto-correlation estimator for a non-negative weight.

\section{Methodology}\label{sec:method}

\subsection{Power spectrum analysis pipeline}\label{subsec:EFT}

Within the framework of the Effective Field Theory of Large-Scale Structure (EFTofLSS)~\citep{Perko:2016puo}, the redshift-space galaxy power spectrum is modeled as
\begin{compactwidetext}
\begin{align}
    P_{g}(z,k,\mu) &= Z_1(z,\mu)^2 P_{11}(z,k) \label{eq:powerspectrum} \\
    &\quad +2\int_{\mathbf{q}}Z_2(z,\mathbf{q},\mathbf{k}-\mathbf{q},\mu)^2 P_{11}(z,|\mathbf{k}-\mathbf{q}|)P_{11}(z,q)
    +6Z_1(z,\mu)P_{11}(z,k)\int_{\mathbf{q}}Z_3(z,\mathbf{q},-\mathbf{q},\mathbf{k},\mu) P_{11}(z,q) \nonumber\\
    &\quad +2Z_1(z,\mu)P_{11}(z,k)\Bigl( c_\mathrm{ct}\frac{k^2}{k_M^2}+ c_{r,1}\mu^2 \frac{k^2}{k_R^2} + c_{r,2}\mu^4 \frac{k^2}{k_R^2}\Bigr) 
    +\frac{1}{\bar{n}_g(z)}\Bigl(c_{\epsilon,0} + c_{\epsilon,\mathrm{mono}}\frac{k^2}{k_M^2} + 3c_{\epsilon,\mathrm{quad}}\bigl(\mu^2-\tfrac{1}{3}\bigr)\frac{k^2}{k_M^2}\Bigr). \nonumber
\end{align}
\end{compactwidetext}
Here $\int_{\mathbf q}\equiv\int d^3q/(2\pi)^3$. The first three terms contain the linear Kaiser contribution and the one-loop corrections, with $P_{11}$ denoting the linear matter power spectrum. The remaining terms are the EFT counterterms and stochastic contributions, parameterized by $\{c_{\rm ct},c_{r,1},c_{r,2},c_{\epsilon,0},c_{\epsilon,\rm mono},c_{\epsilon,\rm quad}\}$, which absorb the effects of unresolved nonlinear physics and discreteness. The scale $k_M^{-1}$ corresponds to the characteristic scale of the spatial derivative expansion, set by the typical extent of the host halo~\citep{Senatore:2014eva}, while $k_R^{-1}$ represents the renormalization scale for velocity products in the redshift-space expansion~\citep{Senatore:2014vja}. Following \citet{Simon:2022csv}, we adopt $k_M = 0.7\,h\,\mathrm{Mpc}^{-1}$ and $k_R = 0.25\,h\,\mathrm{Mpc}^{-1}$ for the eBOSS DR16 QSO analysis.

The functions $Z_n$ in Eq.~\eqref{eq:powerspectrum} are the $n$th-order redshift-space galaxy density kernels~\citep{Perko:2016puo}, given by
\begin{align}\label{eq:redshift_kernels}
    Z_1(z,q_1,\mu) &= K_1(z,q_1) + f(z)\mu_1^2 G_1(q_1) \\
    &= b_1(z) + f(z)\mu_1^2, \nonumber\\
    Z_2(z,q_1,q_2,\mu) &= K_2(z,q_1,q_2)
    + f(z)\mu_{12}^2 G_2(q_1,q_2) \nonumber\\
    &\quad + \frac{1}{2}f(z)\mu q \biggl[
    \frac{\mu_2}{q_2}G_1(q_2)Z_1(z,q_1)+\mathrm{perm.}\biggr],\nonumber \\
    Z_3(z,q_1,q_2,q_3,\mu) &= K_3(z,q_1,q_2,q_3)
    + f(z)\mu_{123}^2 G_3(q_1,q_2,q_3) \nonumber\\
    &\quad + \frac{1}{3}f(z)\mu q \biggl[
    \frac{\mu_3}{q_3}G_1(q_3)Z_2(z,q_1,q_2,\mu_{123}) \nonumber\\
    &\qquad + \frac{\mu_{23}}{q_{23}}G_2(q_2,q_3)Z_1(z,q_1) \nonumber\\
    &\qquad + \mathrm{cyc.}\biggr].\nonumber
\end{align}
In these expressions, $\mu=\mathbf{q}\cdot\hat{\mathbf{z}}/q$, $\mathbf{q}=\mathbf{q}_1+\cdots+\mathbf{q}_n$, $\mu_{i_1\ldots i_n}=\mathbf{q}_{i_1\ldots i_n}\cdot\hat{\mathbf{z}}/q_{i_1\ldots i_n}$, and $\mathbf{q}_{i_1\ldots i_m}=\mathbf{q}_{i_1}+\cdots+\mathbf{q}_{i_m}$. The linear bias $b_1(z)$ is treated as redshift dependent, and $f(z)\equiv d\ln D/d\ln a$ is the linear growth rate. The functions $G_i$ are the standard perturbation-theory velocity kernels, while $K_i$ are the galaxy-density kernels; their explicit expressions are given by Eqs.~(4)--(6) of \citet{Simon:2022csv}.

Observed angles and redshifts are converted into comoving coordinates using a fiducial cosmology. A mismatch between the fiducial and true cosmologies induces an Alcock--Paczynski (AP) distortion in the $(k,\mu)$ plane~\citep{Alcock:1979mp}. The AP-distorted multipoles are given by~\citep{Ballinger:1996cd}
\begin{align} 
    P^\mathrm{AP}_\ell(z,k) &= \frac{2\ell+1}{2q_\perp(z)^2 q_\parallel(z)}\int_{-1}^1 P\bigl[z,k', \mu'\bigr]\,\mathcal{L}_\ell(\mu)\,d\mu, \label{eq:AP-effect}\\
    k'(z,k,\mu) &= \frac{k}{q_\perp(z)}\left[ 1 + \mu^2 \bigl(F(z)^{-2} - 1\bigr) \right]^{1/2},\nonumber\\
    \mu'(z,k,\mu) &= \frac{\mu}{F(z)} \left[ 1 + \mu^2 \bigl(F(z)^{-2} - 1\bigr) \right]^{-1/2}.\nonumber
\end{align}
\begin{align}
    q_\perp(z) &= \frac{D_M(z)}{D_M^\mathrm{fid}(z)},\nonumber \\
    q_\parallel(z) &= \frac{D_H(z)}{D_H^\mathrm{fid}(z)},\nonumber \\
    F(z) &= \frac{q_\parallel(z)}{q_\perp(z)} .\nonumber
\end{align}
Here $D_M(z)$ is the comoving angular-diameter distance and $D_H(z)=c/H(z)$. The scaling parameters are independent of the sound horizon and of the Hubble constant because the shape of the linear power spectrum is varied consistently and distances are expressed in $h^{-1}\,\mathrm{Mpc}$ units \citep[see][]{DAmico:2019fhj,eBOSS:2020yzd}. We use the monopole and quadrupole throughout the analysis, since these two multipoles contain most of the signal-to-noise for the present QSO sample.

The power-spectrum multipoles are computed with \code{Eftpipe}\footnote{\url{https://github.com/zhaoruiyang98/eftpipe}}~\citep{Zhao:2023ebp}, a modified version of \code{PyBird}\footnote{\url{https://github.com/pierrexyz/pybird}}~\citep{DAmico:2020kxu}. The linear power spectrum is obtained from the Boltzmann solver \code{CLASS}\footnote{\url{https://github.com/lesgourg/class_public}}~\citep{Blas:2011rf}, and the one-loop integrals are evaluated with the FFTLog technique~\citep{Simonovic:2017mhp}. We apply IR resummation to account for the damping of BAO features by long-wavelength displacements \citep{Senatore:2014via,Lewandowski:2015ziq,DAmico:2020kxu}. The priors adopted for the cosmological and EFT parameters are listed in Table~\ref{tab:pybird_priors} \citep{Zhao:2023ebp,Maus:2024sbb}.

{\small
\begin{table}[htbp]
\centering
\caption{Priors adopted for the cosmological and EFT parameters. $\mathcal{U}$ and $\mathcal{N}$ denote uniform and Gaussian priors, respectively. For a uniform prior, the two values indicate the lower and upper bounds; for a Gaussian prior, they give the mean and standard deviation. Fixed values are listed as constants.}
\label{tab:pybird_priors}
\setlength{\tabcolsep}{4pt}
\renewcommand{\arraystretch}{1.2}
\begin{tabular}{lclc}
\hline
\multicolumn{2}{c}{Cosmological parameters} & \multicolumn{2}{c}{EFT parameters} \\
\hline
$\Omega_b h^2$ & $\mathcal{N}[0.02237,0.00037]$ & $b_1$ & $\mathcal{U}[0,4]$ \\
$\Omega_{\rm cdm} h^2$ & $\mathcal{U}[0.03,0.7]$ & $b_2$ & $1$ (fixed) \\
$h$ & $\mathcal{U}[0.4,1.0]$ & $b_3$ & $1$ (fixed) \\
$\ln(10^{10}A_s)$ & $\mathcal{U}[0.1,10]$ & $b_4$ & $\mathcal{U}[-15,15]$ \\
$n_s$ & $0.9611$ (fixed) & $c_{\rm ct}$ & $\mathcal{N}[0,4]$ \\
$\Sigma m_{\nu}$  &$0.06\,\mathrm{eV}$(fixed) & $c_{\rm r,1}$ & $\mathcal{N}[0,4]$ \\
$\tau_{\rm reio}$ & $0.055$ (fixed) & $c_{\rm r,2}$ & $0$ (fixed) \\
$w_0$ (CPL) & $\mathcal{U}[-3,1]$ & $c_{\epsilon,0}$ & $\mathcal{N}[0,4]$ \\
$w_a$ (CPL) & $\mathcal{U}[-3,2]$ & $c_{\epsilon,\rm mono}$ & $\mathcal{N}[0,4]$ \\
 & & $c_{\epsilon,\rm quad}$ & $\mathcal{N}[0,4]$ \\
\hline
\end{tabular}
\end{table}
}

\subsection{The Karhunen--Lo\`eve compression}\label{subsec:KL}

To retain the tomographic information encoded in the redshift evolution of the power-spectrum multipoles while avoiding a high-dimensional tomographic data vector, we formulate the weighting scheme as a Karhunen--Lo\`eve (KL), or Fisher-optimal, compression problem \citep{Tegmark1997,Zhu:2014ica}. Let $\mathbf{P}$ denote the uncompressed data vector, composed of $P_\ell(k,z)$ measurements over redshift intervals, wavenumber bins, and the multipoles $\ell=0,2$. A linear compression is defined by
\begin{equation}
    \mathbf{P}_{\rm w}=\mathbf{W}^{\mathrm T}\mathbf{P},\qquad
    \mathbf{C}_{\rm w}=\mathbf{W}^{\mathrm T}\mathbf{C}\mathbf{W},
\end{equation}
where $\mathbf{C}$ is the covariance of the uncompressed vector and $\mathbf{W}$ is the weight matrix. The Fisher matrix for parameters $\boldsymbol{\theta}=\{\theta_i\}$ is
\begin{equation}
\mathbf{F}=\mathbf{D}^{\mathrm T}\mathbf{C}^{-1}\mathbf{D},\qquad
\mathbf{D}=\left(
\frac{\partial\mathbf{P}}{\partial\theta_1},
\frac{\partial\mathbf{P}}{\partial\theta_2},
\ldots,
\frac{\partial\mathbf{P}}{\partial\theta_N}
\right).
\end{equation}
After compression the Fisher matrix becomes
\begin{equation}
\mathbf{F}_{\rm w}=\mathbf{D}_{\rm w}^{\mathrm T}\mathbf{C}_{\rm w}^{-1}\mathbf{D}_{\rm w},\qquad
\mathbf{D}_{\rm w}=\mathbf{W}^{\mathrm T}\mathbf{D}.
\end{equation}
For a Gaussian likelihood with locally linear parameter dependence, the Fisher information is preserved by the MOPED/KL choice,
\begin{equation}\label{eq:KL weight}
    \mathbf{W}=\mathbf{C}^{-1}\mathbf{D}.
\end{equation}

In practice the assumptions of Gaussianity and local linearity are approximate, so the weights should be regarded as a near-optimal compression rather than an exact sufficient statistic. This is sufficient for the present application because the weights are used to construct a compact set of observables whose performance is then tested directly on mocks.

\section{The optimal redshift weights}\label{sec:weights}

\subsection{The data sets}\label{subsec:datasets}

This section summarizes the DR16 QSO catalog and the EZ mock catalogs used for the weighted full-shape analysis.

\subsubsection{The eBOSS DR16 QSO sample}\label{subsubsec:data}

The extended Baryon Oscillation Spectroscopic Survey (eBOSS) \citep{Dawson2016}, part of SDSS-IV \citep{Blanton2017}, was carried out with the 2.5-meter Sloan Foundation Telescope \citep{Gunn2006} at Apache Point Observatory. Quasar target selection is described in \citet{Myers2015} and uses optical SDSS imaging together with mid-infrared photometry from the Wide-field Infrared Survey Explorer (WISE) \citep{Wright2010}. Spectroscopic observations were obtained with the BOSS double-arm spectrographs \citep{Smee2013}.

\begin{figure}
    \centering
    \includegraphics[width=\columnwidth]{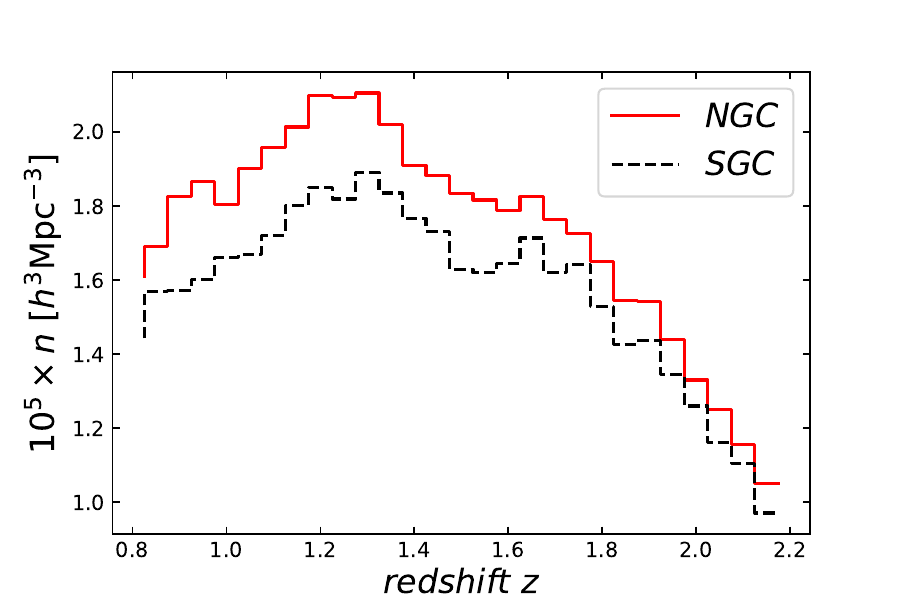}
    \caption{Comoving number density of the DR16 QSO sample in the NGC and SGC regions, shown in redshift bins of width $\Delta z=0.05$.}
    \label{fig_nz}
\end{figure}

We use the eBOSS DR16 quasar clustering catalogs\footnote{\url{https://data.sdss.org/sas/dr16/eboss/lss/catalogs/DR16/}} described in \citet{Ross2020}. The sample covers $0.8<z<2.2$ and contains 343,708 quasars with reliable redshifts over a weighted effective area of $4,699\,\mathrm{deg}^2$, split into $2,860\,\mathrm{deg}^2$ in the North Galactic Cap (NGC) and $1,839\,\mathrm{deg}^2$ in the South Galactic Cap (SGC). The comoving number density is shown in Fig.~\ref{fig_nz}.

The baseline weight assigned to each quasar is
\begin{equation}
    w_{\rm base}=w_{\rm sys}\,w_{\rm cp}\,w_{\rm noz}\,w_{\rm FKP},
    \label{eq:base_weight}
\end{equation}
where $w_{\rm sys}$ corrects for imaging systematics such as seeing, Galactic extinction, and sky background; $w_{\rm cp}$ and $w_{\rm noz}$ correct for fiber collisions and redshift failures, respectively \citep{Ross2020}; and $w_{\rm FKP}=(1+\bar n P_0)^{-1}$ is the Feldman--Kaiser--Peacock weight \citep{Feldman1994}, with $P_0=6000\,h^{-3}\,\mathrm{Mpc}^3$. The optimal redshift weights described below are applied on top of this baseline weight in the power-spectrum estimator.

\subsubsection{Simulated mock catalogs}\label{subsec:mock}

The covariance matrix and pipeline validation require mock catalogs that reproduce both the clustering and the survey selection of the DR16 QSO sample. We use 1000 Extended Zel'dovich (EZ) light-cone mocks \citep{Chuang2015}, whose construction, calibration, and validation are described in \citet{eBOSS:2020wwo}.

The EZ mocks were generated in a flat $\Lambda$CDM cosmology based on Planck2013, with $\Omega_m=0.307115$, $\Omega_b=0.048206$, $h=0.6777$, $\sigma_8=0.8225$, and $n_s=0.9611$ \citep{eBOSS:2020wwo}. For the conversion of observed redshifts to comoving distances we adopt the eBOSS full-shape fiducial cosmology, $\Omega_{\rm m}=0.31$ and $h=0.676$ \citep{eBOSS:2020yzd}. The difference between these two cosmologies is negligible for the purposes of the present analysis. The light-cone mocks, built by stacking simulation boxes of side length $5\,h^{-1}\,\mathrm{Gpc}$ at seven redshift snapshots, retain the redshift evolution required for testing the optimal weights.

\subsection{Construction of the optimal redshift weights}\label{subsubsec:weight}

We construct weights for two cosmological models. In $\Lambda$CDM, the target parameters are $\boldsymbol{\theta}_{\Lambda\rm CDM}=\{h,\Omega_m,\ln(10^{10}A_s)\}$. In the CPL model \citep{Chevallier2001,Linder2003}, the target vector is $\boldsymbol{\theta}_{\rm CPL}=\{h,\Omega_m,\ln(10^{10}A_s),w_0,w_a\}$, with the dark-energy equation of state,
\begin{equation}
    w(z)=w_0+w_a(1-a)=w_0+w_a\frac{z}{1+z}.
    \label{eq:CPL}
\end{equation}

The weights are computed from Eq.~\eqref{eq:KL weight}. For this calculation the redshift-binned covariance is evaluated with the Gaussian analytic expression \citep{Taruya:2010mx},
\begin{align}
    \mathrm{Cov}_{\ell\ell'}(z,k) &= \frac{(2\ell+1)(2\ell'+1)}{2}
    \frac{4\pi^2}{k^2\Delta k\,\Delta V(z)} \nonumber\\
    &\quad\times \int_{-1}^{1} d\mu\,\mathcal{L}_{\ell}(\mu)\mathcal{L}_{\ell'}(\mu) \nonumber\\
    &\quad\times
    \left[P_g(z,k,\mu)+\frac{1}{\bar{n}_{g}(z)}\right]^2.
\end{align}
The model power spectrum $P_g(z,k,\mu)$ is predicted by the EFT pipeline described in Sec.~\ref{subsec:EFT}. The derivative matrix is evaluated numerically as
\begin{align}
    \frac{\partial P_{\ell}(k,z)}{\partial \theta_i}
    &\simeq \frac{1}{2\Delta\theta_i}
    \left[P_{\ell}(k,z)|_{\theta_i+\Delta\theta_i}
    -P_{\ell}(k,z)|_{\theta_i-\Delta\theta_i}\right].
\end{align}
The derivatives are taken around the EZ mock cosmology, using the best-fitting EFT parameters obtained from fitting the mean NGC EZ mock spectrum. The redshift dependence of the linear bias follows \citet{eBOSS:2017ozs},
\begin{equation} \label{eq:bias}
    b_1(z)=0.278\left[(1+z)^2 -6.565\right]+2.393.
\end{equation}
The baseline values used for the weight calculation are summarized in Table~\ref{tab:results}.

{\small
\begin{table}[htbp]
\centering
\caption{Baseline parameter values adopted for the computation of the optimal redshift weights. The EFT parameters are obtained from the mean EZ mock fit used as the expansion point for the numerical derivatives.}
\label{tab:results}
\setlength{\tabcolsep}{4pt}
\renewcommand{\arraystretch}{1.2}
\begin{tabular}{lccc}
\hline
\multicolumn{2}{c}{Cosmological parameters} & \multicolumn{2}{c}{EFT parameters} \\
\hline
$\Omega_b h^2$     & 0.022161 & $b_1$    & Eq.~\eqref{eq:bias} \\
$\Omega_{\rm cdm} h^2$ & 0.11889  & $b_2$    & 1.0\\
$h$                & 0.6777   & $b_3$    & 1.0\\
$\ln(10^{10}A_s)$  & 3.0973   & $b_4$    & 1.4046\\
$n_s$              & 0.9611   & $c_{\rm ct}$ & $-2.9743$\\
$\Sigma m_{\nu}$   & $0.06\,\mathrm{eV}$ & $c_{r,1}$ & $-2.1177$\\
$\tau_{\rm reio}$  & 0.055    & $c_{r,2}$ & $0.0$\\
$w_0$              & $-1.0$   & $c_{\epsilon,0}$ & $-0.0039451$\\
$w_a$              & $0.0$    & $c_{\epsilon,\mathrm{mono}}$ & $0.025473$\\
$n_g(z)$           & Fig.~\ref{fig_nz} & $c_{\epsilon,\mathrm{quad}}$ & $0.094762$\\
\hline
\end{tabular}
\end{table}
}

We adopt $\Delta\theta=0.001$ for the finite differences and have verified that the resulting weights are stable against the step size: using a step equal to $1\%$ of the fiducial value changes the weight functions by less than $2\%$. The weights also depend only weakly on Fourier scale over $0.05\leq k\leq0.25\,h\,\mathrm{Mpc}^{-1}$, so we use $k=0.1\,h\,\mathrm{Mpc}^{-1}$ as the reference scale, following the redshift-weighting literature \citep{eBOSS:2018yfg}.

\begin{figure}
    \centering
    \includegraphics[width=\columnwidth]{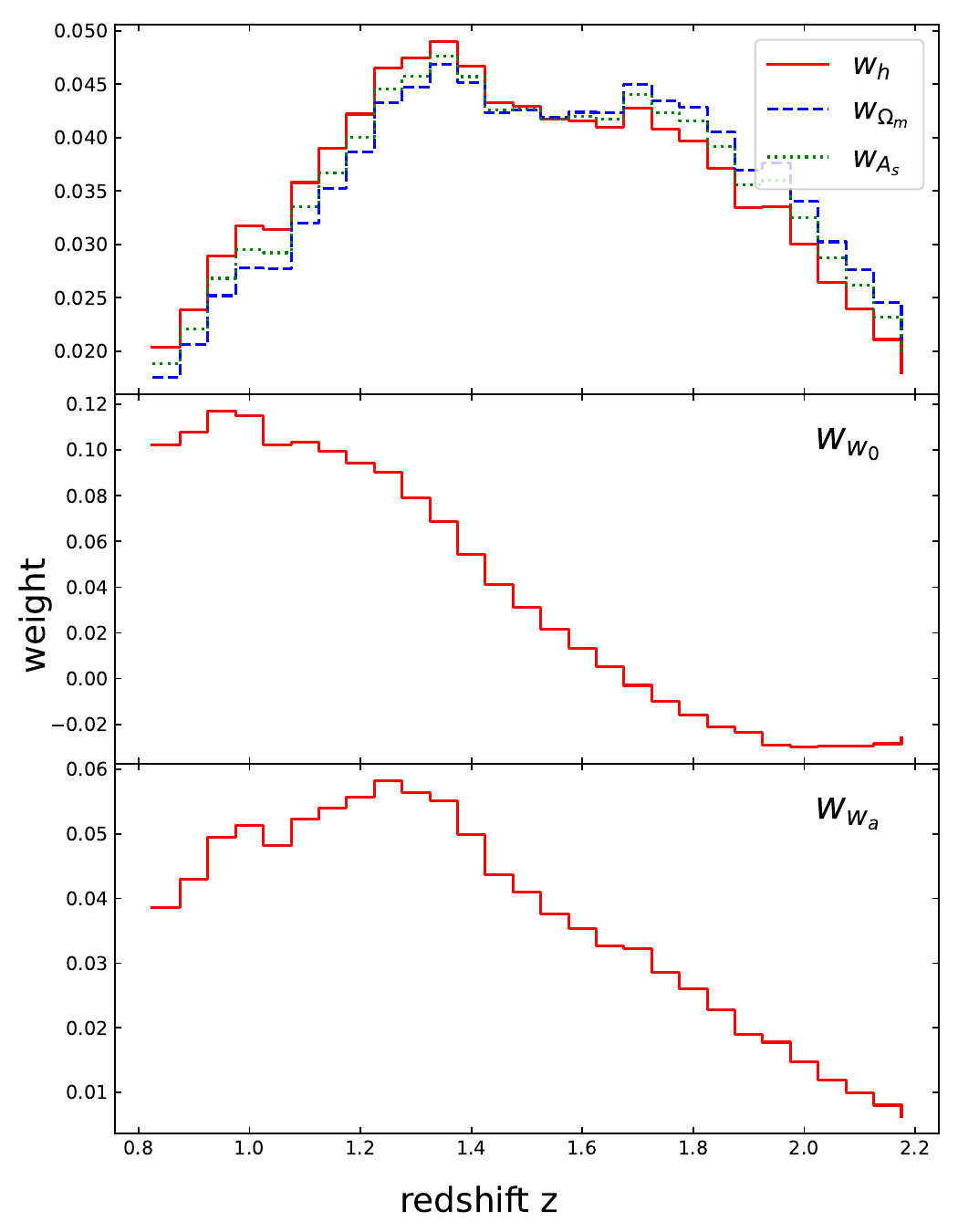}
    \caption{Optimal redshift weights for the power-spectrum monopole. The parameter targeted by each weight is indicated in the legend. Each curve is normalized to unit sum over the redshift bins in $0.8<z<2.2$.}
    \label{fig_w_all}
\end{figure}

Figure~\ref{fig_w_all} shows the monopole weights, normalized to unit sum over the redshift bins. The weights targeting $h$, $\Omega_m$, and $A_s$ have very similar shapes, while $W_{w_0}$ and $W_{w_a}$ emphasize lower redshifts and therefore carry more information about dark-energy evolution. The quadrupole weights show the same qualitative behavior and are not plotted separately.

\subsection{Measurement of the weighted galaxy power spectrum}
\label{subsec:pk_measurement}

We measure the power-spectrum multipoles with the FFT-based Yamamoto estimator \citep{Yamamoto2006,Bianchi:2015oia,Hand:2017irw}, as implemented in \texttt{pypower}.\footnote{\url{https://github.com/cosmodesi/pypower}} The weight shapes in Fig.~\ref{fig_w_all} are reflected in the effective redshifts listed in Table~\ref{tab:zeff}. The weights $W_h$, $W_{\Omega_m}$, and $W_{A_s}$ yield effective redshifts close to the standard value $z_{\rm eff}\simeq1.52$, whereas $W_{w_0}$ and $W_{w_a}$ move the effective response to lower redshift. This is why the weights are expected to be most useful for the CPL model, especially for $w_a$.

\begin{table}[htbp]
\centering
\caption{Effective redshifts of the eBOSS DR16 QSO sample and the mean of the 1000 EZ mock catalogs. Values are shown for the standard FKP measurement and for the representative optimal weights targeting $h$, $\Omega_m$, $A_s$, $w_0$, and $w_a$.}
\label{tab:zeff}
\renewcommand{\arraystretch}{1.2}
\setlength{\tabcolsep}{6pt}
\begin{tabular}{lcccc}
\toprule
\multirow{2}{*}{Weighting scheme} & \multicolumn{2}{c}{Data} & \multicolumn{2}{c}{EZ mock} \\
 & NGC & SGC & NGC & SGC \\
\midrule
No $z$ weight & 1.52 & 1.52 & 1.52 & 1.53 \\
$W_h$, $W_{\Omega_{\rm m}}$, $W_{A_{\rm s}}$ & 1.53 & 1.54 & 1.53 & 1.54 \\
$W_{w_0}$ & 0.94 & 0.92 & 0.92 & 0.91 \\
$W_{w_a}$ & 1.37 & 1.37 & 1.37 & 1.37 \\
\bottomrule
\end{tabular}
\end{table}

Because $W_h$, $W_{\Omega_m}$, and $W_{A_s}$ are nearly degenerate in redshift dependence, we keep only $W_h$ as their representative in the measured data vector. For the CPL model we use $W_h$, $W_{w_0}$, and $W_{w_a}$. This choice captures the relevant redshift-evolution information while avoiding redundant observables in the covariance matrix.

The weighted spectra are measured with a cross-correlation estimator. This is essential for sign-changing weights such as $W_{w_0}$ and also allows the same measurement convention to be used for all the weighted spectra. For a target redshift weight $W_z$, we construct two FKP fields with weights,
\begin{equation}
    w^{(1)}=w_{\rm base}W_z,\qquad
    w^{(2)}=w_{\rm base}.
    \label{eq:cross_weights}
\end{equation}
The cross-power multipole estimator is
\begin{align}
    \hat{P}^{(12)}_{\ell}(k) &=\frac{2\ell+1}{A_{12}} \int \frac{d \Omega_k}{4 \pi} F^{(1)}_0(\mathbf{k}) F^{(2)}_{\ell}(-\mathbf{k})-P^{(12)}_{\ell,{\rm shot}}, \label{eq:cross_estimator}\\
    F^{(j)}_{\ell}(\mathbf{k}) &=\int d \mathbf{r}\, F^{(j)}(\mathbf{r}) \mathcal{L}_{\ell}(\hat{\mathbf{k}} \cdot \hat{\mathbf{r}}) e^{i \mathbf{k} \cdot \mathbf{r}},\qquad j=1,2 .\nonumber
\end{align}
Here $F^{(j)}(\mathbf{r})=n^{(j)}_D(\mathbf{r})-\alpha^{(j)}_R n^{(j)}_R(\mathbf{r})$ is the FKP field, $n^{(j)}_D$ and $n^{(j)}_R$ are the weighted galaxy and random densities painted onto the FFT grid, and $\alpha^{(j)}_R=\sum_i w^{(j)}_{D,i}/\sum_i w^{(j)}_{R,i}$. The shot-noise contribution is non-zero only for the monopole:
\begin{equation}
P^{(12)}_{0,{\rm shot}}=\frac{1}{A_{12}}\left[\sum_{i=1}^{N_D}w^{(1)}_{D,i}w^{(2)}_{D,i}+\alpha_R^{(1)}\alpha_R^{(2)}\sum_{i=1}^{N_R}w^{(1)}_{R,i}w^{(2)}_{R,i}\right].
\label{eq:cross_shot}
\end{equation}
The cross-normalization is
\begin{align}\label{eq:cross_norm}
A_{12} &= \frac{1}{2}\left[\alpha_R^{(1)}\int d\mathbf{r}\,
    n_R^{(1)}(\mathbf{r})n_D^{(2)}(\mathbf{r}) \right. \\
&\quad\left. +\alpha_R^{(2)}\int d\mathbf{r}\,
    n_R^{(2)}(\mathbf{r})n_D^{(1)}(\mathbf{r})\right].\nonumber
\end{align}

For a non-negative weight, the conventional auto-spectrum implementation would weight both fields by $w_{\rm base}\sqrt{W_z}$, giving the same pair weight at leading order. The small finite-volume difference between this conventional auto estimator and the cross estimator arises from the asymmetric field construction and from the corresponding normalization and shot-noise terms. We quantify this consistency check in Appendix~\ref{app:cross_auto}; the main analysis uses the cross estimator defined above.

\begin{figure*}
    \centering
    \includegraphics[width=0.88\textwidth]{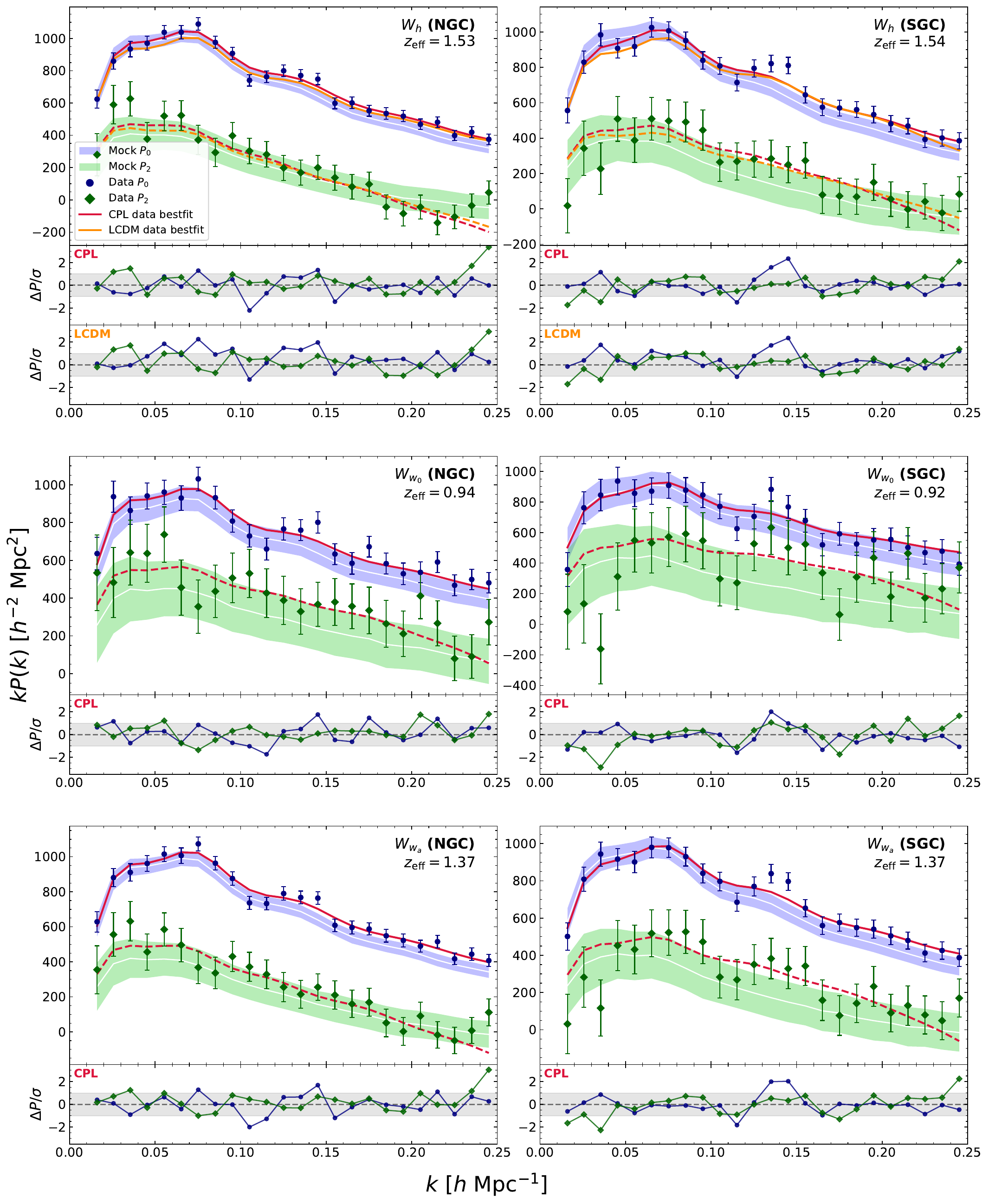}
    \caption{Weighted power-spectrum monopole and quadrupole measured from the DR16 QSO sample, compared with the mock expectation and the best-fitting models. The left and right columns show NGC and SGC; the rows correspond to $W_h$, $W_{w_0}$, and $W_{w_a}$. Blue circles and green diamonds denote the data measurements of $P_0$ and $P_2$, while the blue and green shaded bands show the mean and $1\sigma$ scatter of the same measurements from the 1000 EZ mocks. Red and orange curves show the best-fitting CPL and $\Lambda$CDM predictions, respectively, with solid curves for $P_0$ and dashed curves for $P_2$. All spectra in the main panels are multiplied by $k$ for display. The narrow lower panels show the normalized data-model residuals, $\Delta P_\ell/\sigma_\ell=[P_\ell^{\rm data}(k)-P_\ell^{\rm bf}(k)]/\sigma_\ell(k)$, where $\sigma_\ell(k)$ is the diagonal $1\sigma$ error bar of the corresponding data point from the mock covariance. The grey band therefore marks residuals within one statistical error bar of the best-fitting model.}
    \label{fig:pk_results}
\end{figure*}

Following this procedure, we measure the redshift-weighted spectra for the data and for all 1000 EZ mock catalogs. Figure~\ref{fig:pk_results} shows the measured monopole and quadrupole for the representative weights $W_h$, $W_{w_0}$, and $W_{w_a}$, separately for the NGC and SGC footprints.

Cosmological parameters are inferred with a Gaussian likelihood,
\begin{equation}\label{chi2}
\chi^2=\left(\mathbf{P}^{\rm data}-\mathbf{P}^{\rm win}\right)^{\!\mathrm T}\mathbf{C}^{-1}\left(\mathbf{P}^{\rm data}-\mathbf{P}^{\rm win}\right),
\end{equation}
where $\mathbf{P}^{\rm data}$ is the measured weighted data vector, $\mathbf{C}$ is the covariance estimated from the mocks with the same weights, and $\mathbf{P}^{\rm win}$ is the theory vector after convolution with the survey window. We apply the window convolution directly in Fourier space,
\begin{equation}
    P^{\rm win}_{\ell}(k_i)=\sum_{\ell'}\sum_j \mathcal{W}_{\ell\ell'}(k_i,k'_j)P^{\rm th}_{\ell'}(k'_j),
    \label{eq:window_convolution}
\end{equation}
with $\ell=0,2$ for the measured vector and $\ell'=0,2,4$ retained in the convolution. The Fourier-space window matrix is computed from the configuration-space window multipoles as \citep{Wilson:2015lup,Beutler:2018vpe,DAmico:2019fhj,Zhao:2023ebp}
\begin{align}\label{eq:window_W}
    \mathcal{W}_{\ell\ell'}(k,k') &= \frac{2}{\pi}(-i)^\ell i^{\ell'}k'^2
    \int ds\,s^2 j_\ell(ks)Q_{\ell\ell'}(s)j_{\ell'}(k's), \\
    Q_{\ell\ell'}(s) &= \sum_{\ell''} C_{\ell\ell'\ell''}\,Q_{\ell''}(s), \nonumber\\
    Q_{\ell}(s) &\propto (2\ell+1)\int\frac{d^2\hat{s}}{4\pi}\int d^3x\,
    \bar{n}(\mathbf{x})\bar{n}(\mathbf{x}+\mathbf{s}) \nonumber\\
    &\quad\times
    \mathcal{L}_\ell(\hat{\mathbf{s}}\cdot\hat{\boldsymbol{\eta}}), \nonumber
\end{align}
where $j_\ell$ is the $\ell$th spherical Bessel function, $\bar n$ is the survey selection function, and $Q_\ell(s)$ is the $\ell$th multipole of the survey-window correlation function. The coupling coefficients used for the monopole and quadrupole are
\begin{equation}
    \begin{aligned}
        C_{0,\ell',\ell''} &=
        \begin{bmatrix}
            1 & 0 & 0 \\
            0 & \frac{1}{5} & 0 \\
            0 & 0 & \frac{1}{9}
        \end{bmatrix}_{\!\ell',\ell''},
        &
        C_{2,\ell',\ell''} &=
        \begin{bmatrix}
            0 & 1 & 0 \\
            1 & \frac{2}{7} & \frac{2}{7} \\
            0 & \frac{2}{7} & \frac{100}{693}
        \end{bmatrix}_{\!\ell',\ell''}.
    \end{aligned}
    \label{eq:C}
\end{equation}

The window multipoles are measured with \texttt{pypower}, following \citet{Beutler2021}, and are recomputed for each redshift weight and Galactic cap. Wide-angle terms are neglected in the baseline convolution because the QSO sample lies at high redshift, $0.8<z<2.2$, and the corresponding corrections are subdominant for the present precision. Since the fitted data vector contains only $P_0$ and $P_2$, retaining theory multipoles up to $\ell'=4$ is sufficient for the window mixing.

\begin{figure}
    \centering
    \includegraphics[width=\columnwidth]{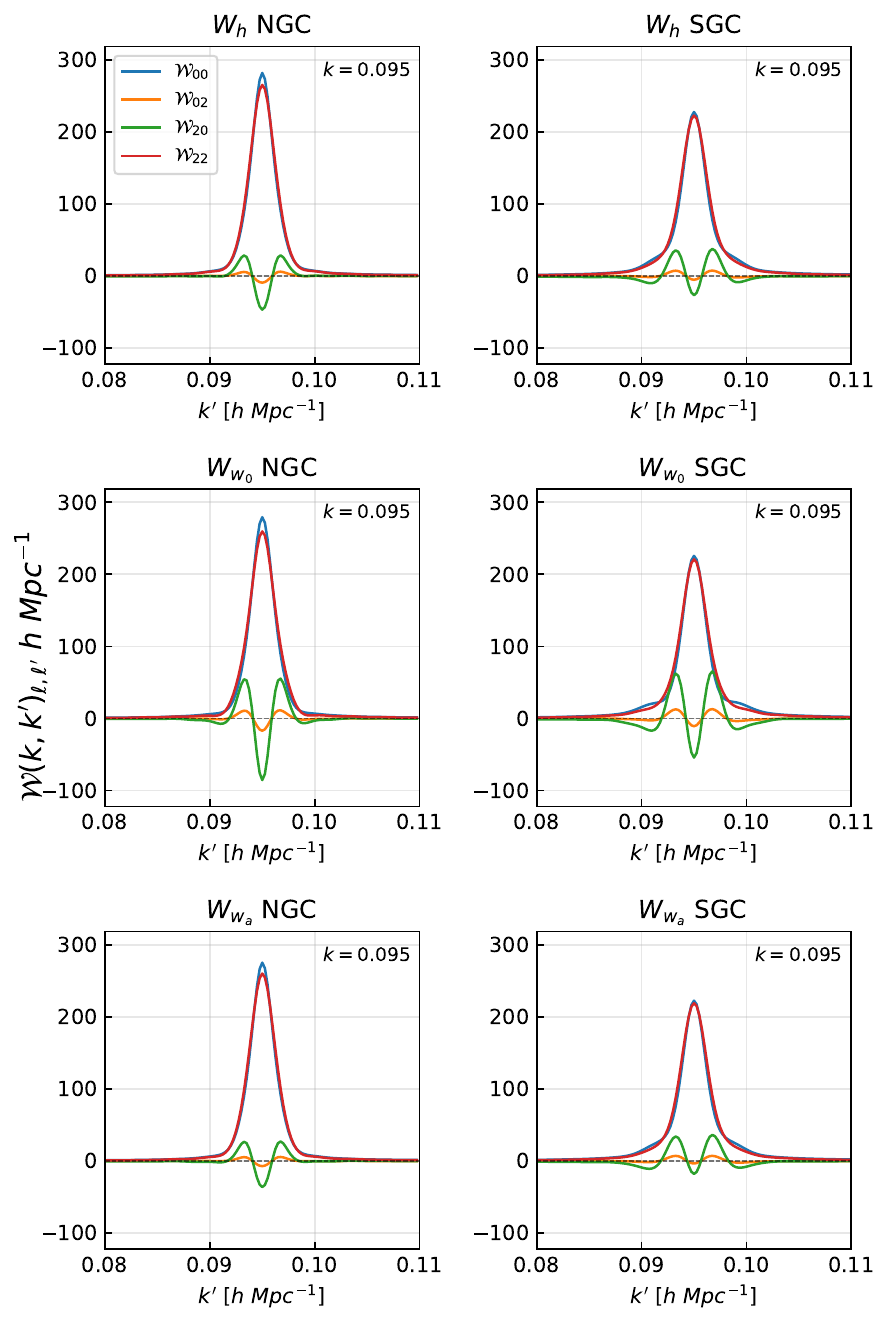}
    \caption{Fourier-space survey-window kernels $\mathcal{W}_{\ell\ell'}(k,k')$ evaluated at $k=0.095\,h\,\mathrm{Mpc}^{-1}$. The left and right columns show NGC and SGC, and the rows correspond to $W_h$, $W_{w_0}$, and $W_{w_a}$. The curves show the multipole mixing entering Eq.~\eqref{eq:window_convolution}; the localization of the kernels around $k'=k$ illustrates that the survey mask acts as a controlled smoothing of the theoretical spectra.}
    \label{fig:window}
\end{figure}

Figure~\ref{fig:window} displays the Fourier-space kernels $\mathcal{W}_{\ell\ell'}(k,k')$ at $k\simeq0.095\,h\,\mathrm{Mpc}^{-1}$ for the three representative weights and both Galactic caps. The kernels are localized around $k'=k$ and vary smoothly with $k'$, which provides a direct numerical check of the Fourier-space convolution used in the likelihood.

The covariance matrix is estimated from the 1000 EZ mocks after applying the identical redshift weights, cross-power measurement, and window-convolved theory comparison used for the data. Since the inverse covariance estimated from a finite number of mocks is biased, we apply the Hartlap correction \citep{Hartlap2007},
\begin{equation}
    \mathbf{C}^{-1}_{\rm debiased}=\frac{N_{\rm mock}-N_{\rm bin}-2}{N_{\rm mock}-1}\,\mathbf{C}^{-1},
\end{equation}
where $N_{\rm mock}=1000$ and $N_{\rm bin}$ is the length of the data vector for the corresponding fit.

\section{Results}\label{sec:result}

\subsection{Analysis for $\Lambda$CDM based on optimal redshift weights}\label{subsec:LCDMr}

We first apply the method to $\Lambda$CDM. In this model the target parameters do not generate a strong differential response across the QSO redshift interval, so the weights mainly test whether the measurement pipeline is unbiased. The fits use the full-shape monopole and quadrupole up to $k_{\rm max}=0.24\,h\,\mathrm{Mpc}^{-1}$ \citep{Simon:2022csv} and are performed both on the mean of the 1000 EZ mocks and on the DR16 QSO data.

\begin{figure}
\centering
\includegraphics[width=\columnwidth]{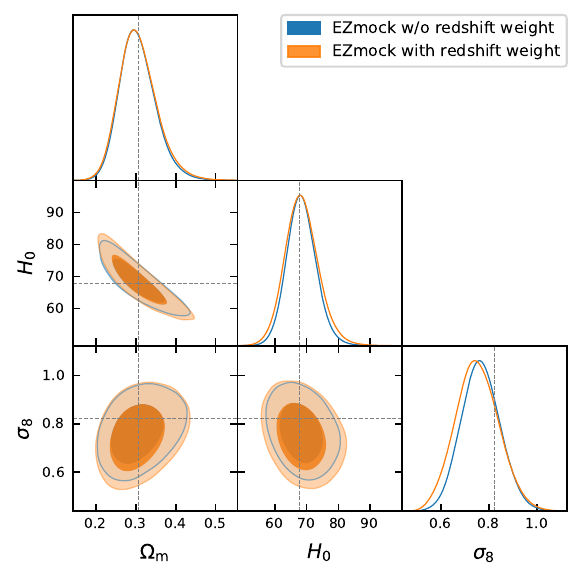}
\caption{Posterior distributions from the mean EZ mock power spectrum in the $\Lambda$CDM model, comparing the standard and redshift-weighted analyses. The fit uses $k_{\rm max}=0.24\,h\,\mathrm{Mpc}^{-1}$. Contours show the 68\% and 95\% credible regions, crosses mark best-fit values, and dashed lines indicate the mock truth.}
\label{fig:LCDM_mock_triangle}
\end{figure}

Figure~\ref{fig:LCDM_mock_triangle} shows the constraints from the mean EZ mock spectrum. The weighted and standard contours are nearly superimposed and recover the fiducial cosmology. This behavior follows directly from the weights: $W_h$, $W_{\Omega_m}$, and $W_{A_s}$ have effective redshifts close to the standard FKP measurement, so little extra information is expected in $\Lambda$CDM.

\begin{figure}
\centering
\includegraphics[width=\columnwidth]{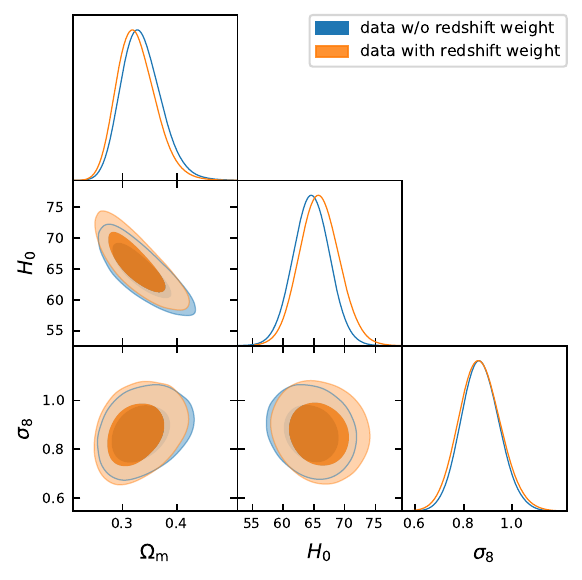}
\caption{Posterior distributions from the eBOSS DR16 QSO data in the $\Lambda$CDM model, comparing the standard and redshift-weighted analyses. The fit uses $k_{\rm max}=0.24\,h\,\mathrm{Mpc}^{-1}$; contours show the 68\% and 95\% credible regions and crosses mark best-fit values.}
\label{fig:LCDM_data_triangle}
\end{figure}

The DR16 QSO data show the same pattern, as displayed in Fig.~\ref{fig:LCDM_data_triangle}. The weighted and standard posteriors are statistically consistent, and the best-fitting $\Lambda$CDM curves in Fig.~\ref{fig:pk_results} describe the weighted monopole and quadrupole measurements within the mock errors. We therefore use the $\Lambda$CDM case primarily as a null test: adding the redshift-weighted observables does not shift the result when no significant gain is expected.

\subsection{Analysis for the CPL model based on optimal redshift weights}\label{subsec:CPLr}

We next consider the CPL model, where $w(z)=w_0+w_a z/(1+z)$ gives the clustering signal an explicit redshift dependence through the background expansion and growth history. The low-redshift response of $W_{w_0}$ and the distinct shape of $W_{w_a}$ are therefore expected to carry information that is partly averaged out by a single effective-redshift measurement.

\begin{figure}
\centering
\includegraphics[width=\columnwidth]{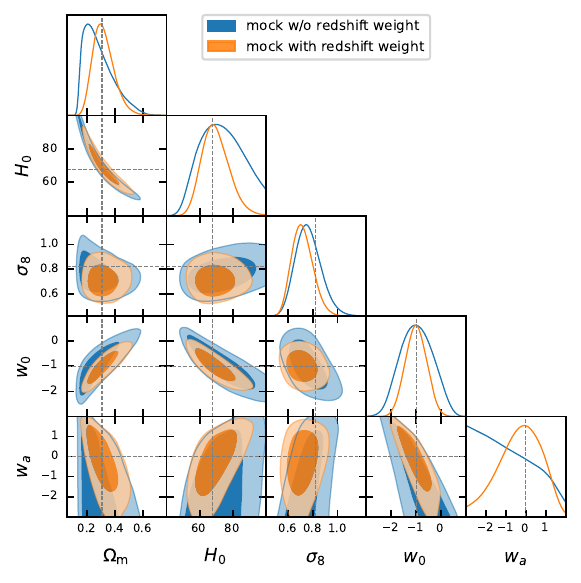}
\caption{Posterior distributions from the mean EZ mock power spectrum in the CPL model, comparing the standard and redshift-weighted analyses. The fit uses $k_{\rm max}=0.24\,h\,\mathrm{Mpc}^{-1}$. Contours show the 68\% and 95\% credible regions, crosses mark best-fit values, and dashed lines indicate the mock truth.}
\label{fig:CPL_mock_triangle}
\end{figure}

The mock test confirms this expectation. Figure~\ref{fig:CPL_mock_triangle} and Table~\ref{tab:CPL_mock} show that the weighted analysis improves the precision on $H_0$, $w_0$, $\Omega_m$, and $\sigma_8$ by $46.7\%$, $33.3\%$, $22.5\%$, and $14.7\%$, respectively. The clearest change is in $w_a$: the standard fit gives a one-sided marginalized constraint, whereas the weighted fit gives a bounded posterior compatible with the fiducial value. This is precisely the role of the weights: to expose the redshift response of the light cone rather than to duplicate the standard measurement.

\begin{table}[t]
\centering
\caption{CPL parameter constraints from the stacked EZ mock catalogs, comparing the standard and redshift-weighted analyses. The improvement is defined as $(\bar\sigma_{\rm std}-\bar\sigma_{\rm w})/\bar\sigma_{\rm std}$, using the average of the upper and lower $1\sigma$ errors for asymmetric posteriors. A value is not quoted for $w_a$ because the standard analysis yields only an upper limit.}
\label{tab:CPL_mock}
\setlength{\tabcolsep}{4pt}
\renewcommand{\arraystretch}{1.2}
\begin{tabular}{lccc}
\toprule
Parameter & Standard & Weighted & Improvement \\
\midrule
$\Omega_m$ & $0.286^{+0.047}_{-0.14}$  & $0.319^{+0.057}_{-0.088}$ & $22.5\%$ \\
$H_0$      & $73^{+10}_{-20}$           & $68.5^{+6.8}_{-9.2}$      & $46.7\%$ \\
$\sigma_8$ & $0.765^{+0.087}_{-0.11}$   & $0.765^{+0.073}_{-0.095}$ & $14.7\%$ \\
$w_0$      & $-0.97 \pm 0.63$           & $-0.998 \pm 0.42$         & $33.3\%$ \\
$w_a$      & ${<\,-0.281}$              & $-0.30^{+1.3}_{-0.88}$    & --- \\
\bottomrule
\end{tabular}
\end{table}

\begin{figure}
\centering
\includegraphics[width=\columnwidth]{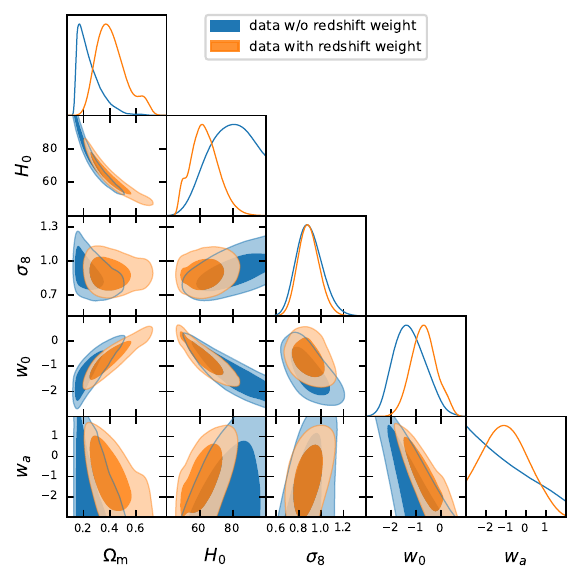}
\caption{Posterior distributions from the eBOSS DR16 QSO data in the CPL model, comparing the standard and redshift-weighted analyses. The fit uses $k_{\rm max}=0.24\,h\,\mathrm{Mpc}^{-1}$; contours show the 68\% and 95\% credible regions and crosses mark best-fit values.}
\label{fig:CPL_data_triangle}
\end{figure}

The DR16 QSO data show the same qualitative behavior. Figure~\ref{fig:CPL_data_triangle} and Table~\ref{tab:CPL_data} show that the weighted analysis reduces the uncertainties on $H_0$, $\sigma_8$, and $w_0$ by $43.3\%$, $19.7\%$, and $20.5\%$, respectively. For $w_a$, the standard fit gives only the one-sided 68\% constraint $w_a<-0.498$, while the weighted fit yields $w_a=-0.98^{+1.0}_{-1.3}$.

The gain is not uniform across all marginalized parameters. The $\Omega_m$ interval broadens in the weighted CPL fit, indicating that the added redshift-evolution information changes the degeneracy directions rather than simply shrinking every one-dimensional error bar. The key result is that the redshift-weighted analysis converts the poorly bounded $w_a$ posterior into a closed constraint, while remaining consistent with $\Lambda$CDM within the reported uncertainties.

\begin{table}[t]
\centering
\caption{CPL parameter constraints from the eBOSS DR16 QSO data. The redshift-weighted analysis gives a bounded posterior for $w_a$ and reduces the uncertainties on $H_0$, $\sigma_8$, and $w_0$. The negative value in the improvement column for $\Omega_m$ indicates a broader marginalized interval after the CPL degeneracy directions are changed by the weighted observables.}
\label{tab:CPL_data}
\setlength{\tabcolsep}{4pt}
\renewcommand{\arraystretch}{1.2}
\begin{tabular}{lccc}
\toprule
Parameter & Standard & Weighted & Improvement \\
\midrule
$\Omega_m$ & $0.246^{+0.030}_{-0.11}$  & $0.415^{+0.075}_{-0.14}$ & $-53.6\%$ \\
$H_0$      & $73^{+20}_{-10}$           & $68.5^{+8}_{-9}$         & $43.3\%$ \\
$\sigma_8$ & $0.765^{+0.099}_{-0.13}$   & $0.765^{+0.084}_{-0.10}$ & $19.7\%$ \\
$w_0$      & $-1.24^{+0.55}_{-0.72}$    & $-0.68^{+0.47}_{-0.54}$  & $20.5\%$ \\
$w_a$      & ${<\,-0.498}$              & $-0.98^{+1.0}_{-1.3}$    & --- \\
\bottomrule
\end{tabular}
\end{table}

The CPL results clarify the role of the optimal weights. They do not simply reweight the same effective-redshift measurement; instead, they add the redshift response needed to constrain time-dependent dark-energy parameters. This information is obtained without splitting the QSO catalog into several redshift bins or estimating a correspondingly larger tomographic covariance matrix.

\section{Conclusions and discussion}\label{sec:conclu}

We have carried out a full-shape analysis of the eBOSS DR16 QSO power spectrum using optimal redshift weights. The method compresses the evolution across $0.8<z<2.2$ into a small set of weighted monopole and quadrupole measurements. A cross-correlation estimator is used for all weights, so that sign-changing weights such as $W_{w_0}$ can be treated without changing the measurement convention. The theoretical spectra are convolved with Fourier-space survey-window kernels measured separately for each cap and weighting scheme.

The tests on 1000 EZ light-cone mocks show that the weighted pipeline is stable at the precision of this analysis. In $\Lambda$CDM, the weighted and standard constraints agree for both mocks and data, consistent with the nearly standard effective redshifts of the weights targeting $h$, $\Omega_m$, and $A_s$. This provides a useful test of the estimator, covariance, and window treatment.

The impact appears in the CPL model, where the parameters introduce redshift-dependent changes to the expansion and growth histories. In the stacked mocks, the optimal weights improve the constraints on $H_0$, $w_0$, $\Omega_m$, and $\sigma_8$, and replace the one-sided standard constraint on $w_a$ with a bounded posterior. The DR16 data show the same pattern: the weighted analysis tightens $H_0$, $w_0$, and $\sigma_8$, and gives $w_a=-0.98^{+1.0}_{-1.3}$ rather than an upper limit. The broader marginalized $\Omega_m$ interval in the weighted CPL fit reflects the modified degeneracy structure after adding redshift-evolution information.

Optimal redshift weighting is therefore a compact alternative to a tomographic full-shape analysis for wide-redshift tracer samples. It keeps the QSO catalog as a single light-cone sample, avoids a large tomographic covariance matrix, and extracts the redshift response most relevant to evolving dark-energy models. The same strategy can be applied to wide-redshift spectroscopic surveys such as DESI and Euclid \citep{DESI2016,Laureijs2011}, and can be combined with higher-order statistics, alternative compression schemes, or multi-tracer analyses \citep{Gualdi:2019ybt,Gualdi:2018pyw,Zhao:2023ebp}.

\begin{acknowledgments}

We thank Ruiyang Zhao for his help with \code{EFTpipe} code and helpful discussions. We thank Cheng Zhao for providing EZ mock catalogs. This work is supported by the National Key R \& D Program of China (2023YFA1607800, 2023YFA1607803, 2025YFA1614103), NSFC grants (12525301, 12273048, 12422301), and by the CAS Project for Young Scientists in Basic Research (No. YSBR-092). YW is also supported by National Key R\&D Program of China No. 2022YFF0503404. GBZ is also supported by science research grants from the China Manned Space Project with No. CMS-CSST-2021-B01, and the New Cornerstone Science Foundation through the XPLORER prize. The authors acknowledge the Beijing Super Cloud Center (\href{https://www.blsc.cn/}{BSCC}) for providing HPC resources that have contributed to the research results reported within this paper.

\end{acknowledgments}

\newpage

\appendix
\setcounter{equation}{0}
\renewcommand{\theequation}{A\arabic{equation}}
\renewcommand{\theHequation}{A.\arabic{equation}}
\setcounter{figure}{0}
\renewcommand{\thefigure}{A\arabic{figure}}
\renewcommand{\theHfigure}{A.\arabic{figure}}

\section{Consistency of the cross- and auto-correlation estimators}
\label{app:cross_auto}

\begin{center}
\refstepcounter{figure}\label{fig:cross_vs_auto_validation}
\includegraphics[width=0.45\textwidth]{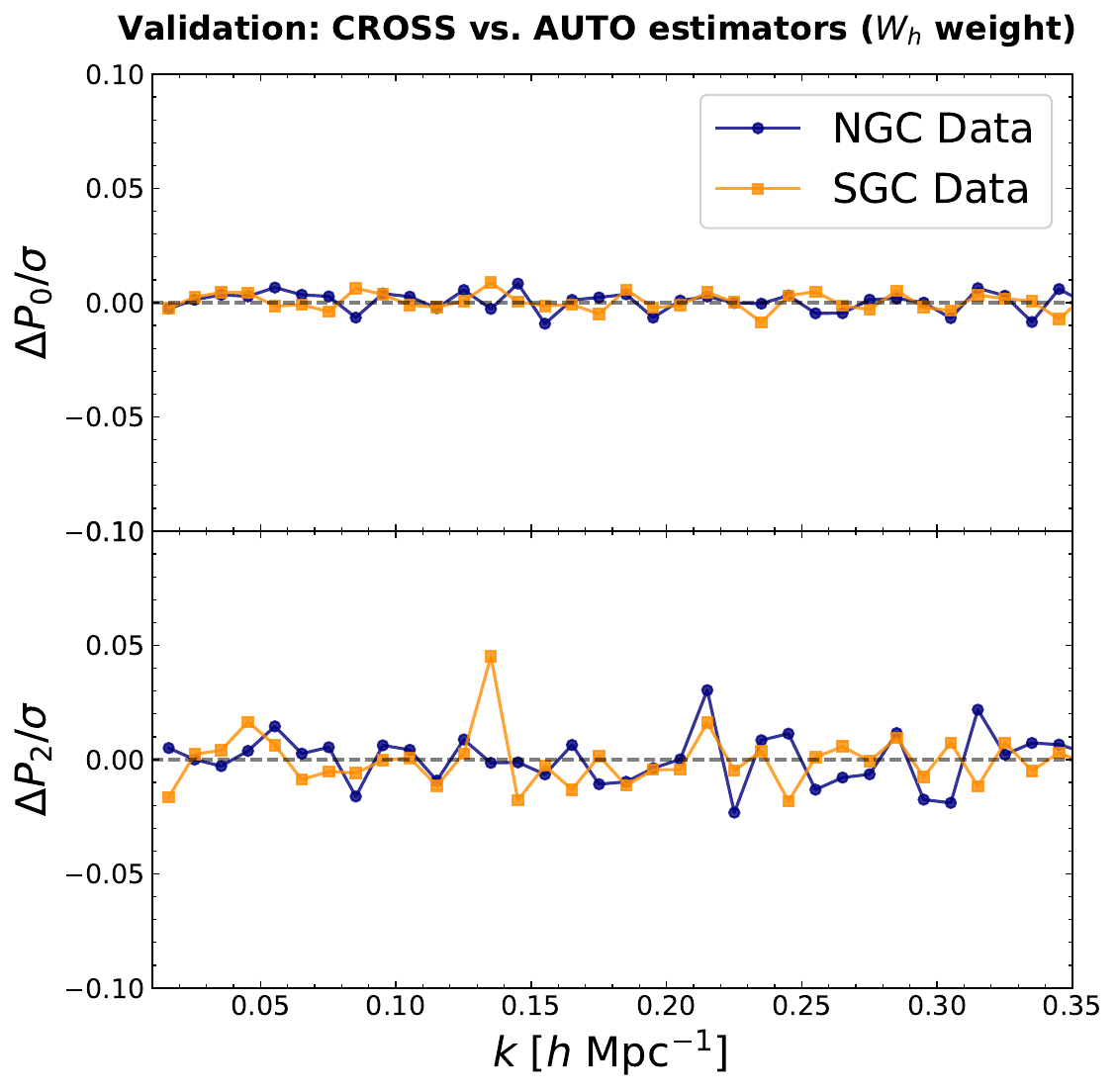}
\par\vspace{0.3em}
\begin{minipage}{0.92\textwidth}
\small\textbf{Figure~\thefigure.} Validation of the cross-correlation estimator using the non-negative $W_h$ weight. The plotted quantity is $\Delta P_\ell/\sigma=(P_{{\rm cross},\ell}-P_{{\rm auto},\ell})/\sigma$, with $\sigma$ taken from the diagonal of the mock covariance for the corresponding $k$ bin and multipole. The residuals are below $0.05\sigma$ for the monopole and quadrupole in both Galactic caps.
\end{minipage}
\end{center}

For completeness, we summarize the estimator check used to validate the cross-correlation measurement. For a non-negative redshift weight $W_z$, the conventional implementation assigns each object the weight $w_{\rm base}\sqrt{W_z}$ and measures an auto-power spectrum. Schematically,
\begin{equation}
    P_{\rm auto}(k)\propto
    \left\langle \sqrt{W_z}\,\delta_{\rm base}(\mathbf{k})\,
    \sqrt{W_z}\,\delta_{\rm base}^*(\mathbf{k})\right\rangle
    \propto w_{\rm base}^2 W_z.
    \label{eq:auto_positive_weight}
\end{equation}
The cross-correlation estimator used in the main text instead correlates fields weighted by $w^{(1)}=w_{\rm base}W_z$ and $w^{(2)}=w_{\rm base}$, giving
\begin{equation}
    P_{\rm cross}(k)\propto
    \left\langle W_z\,\delta_{\rm base}(\mathbf{k})\,
    \delta_{\rm base}^*(\mathbf{k})\right\rangle
    \propto w_{\rm base}^2 W_z.
    \label{eq:cross_positive_weight}
\end{equation}
Thus, for $W_z\geq0$, the two approaches target the same weighted clustering signal. They are not identical finite-volume estimators, because the cross estimator is asymmetric in its two fields and uses the cross-normalization and cross-shot-noise terms in Eqs.~\eqref{eq:cross_shot}--\eqref{eq:cross_norm}. These differences set the expected level of any residual mismatch.

We test this explicitly with the non-negative $W_h$ weight. Figure~\ref{fig:cross_vs_auto_validation} shows the normalized residuals between the cross-correlation and conventional auto-correlation measurements. For both Galactic caps and for both the monopole and quadrupole, the residuals are below $0.05\sigma$. This confirms that the cross estimator reproduces the standard result for positive weights while remaining applicable to sign-changing weights such as $W_{w_0}$.

\bibliography{refs}
\bibliographystyle{aasjournalv7}

\end{document}